\documentclass[letterpaper,english,aps,prb,twocolumn,floatfix,showpacs,amsfonts,amssymb,superscriptaddress]{revtex4}

\usepackage[T1]{fontenc}
\usepackage[latin1]{inputenc}
\usepackage{graphics}
\usepackage{amssymb}
\usepackage{amsmath}
\usepackage{epsfig}

\newcommand{\beq}{\begin{equation}}
\newcommand{\eeq}{\end{equation}}
\newcommand{\bea}{\begin{eqnarray}}
\newcommand{\eea}{\end{eqnarray}}

\newcommand{\rmd}{{\rm d}}

\newcommand{\pinf}{P_{\infty}}
\newcommand{\idt}{\int d \tau}
\newcommand{\idr}{\int d^{2} {\bf r}}

\newcommand{\oi}{{\bf \Omega}_{i}}
\newcommand{\oj}{{\bf \Omega}_{j}}

\newcommand{\kcm}{\frac{k_{F}^{2}c^{2}}{m_{\beta}^{2}}}
\newcommand{\nib}{{\bf n}_{i}}
\newcommand{\njb}{{\bf n}_{j}}
\newcommand{\li}{{\bf l}_{i}}

\newcommand{\nb}{{\bf n}}
\newcommand{\lb}{{\bf l}}
\newcommand{\qb}{{\bf q}}
\newcommand{\on}{\omega_{n}}
\newcommand{\kb}{{\bf k}}
\newcommand{\eabc}{\epsilon_{\alpha \beta \gamma}}
\newcommand{\trace}{\frac{1}{\beta} \sum_{\omega_{n}} \int \frac{d^{2}{\bf k}}{(2\pi)^{2}}}
\newcommand{\ob}{{\bf \Omega}}

\newcommand{\cuoo}{CuO$_{2}$}
\newcommand{\lco}{La$_{2}$CuO$_{4}$}
\newcommand{\lasco}{La$_{2-x}$Sr$_{x}$CuO$_{4}$}
\newcommand{\lczo}{La$_{2}$Cu$_{1-z}$Zn$_z$O$_{4}$}
\newcommand{\lasczo}{La$_{2-x}$Sr$_{x}$Cu$_{1-z}$Zn$_z$O$_{4}$}

\begin{document}
\def\tende#1{\,\vtop{\ialign{##\crcr\rightarrowfill\crcr
\noalign{\kern-1pt\nointerlineskip} \hskip3.pt${\scriptstyle
#1}$\hskip3.pt\crcr}}\,}

\title{Competing impurities and reentrant magnetism in
{\lasczo} revisited. \\ The role of the Dzyaloshinskii-Moriya and
XY anisotropies}

\author{L.\ Adamska}

\affiliation{Institute for Theoretical Physics, University of
Utrecht, Leuvenlaan 4, 3584 CE Utrecht, The Netherlands.}

\author{M.\ B.\ Silva Neto}

\affiliation{Institut f\"ur Theoretische Physik, Universit\"at
Stuttgart, Pfaffenwaldring 57, 70550, Stuttgart, Germany.}

\author{C.\ Morais Smith}

\affiliation{Institute for Theoretical Physics, University of
Utrecht, Leuvenlaan 4, 3584 CE Utrecht, The Netherlands.}

\begin{abstract}
We study the order-from-disorder transition and reentrant
magnetism in {\lasczo} within the framework of a long-wavelength
nonlinear sigma model that properly incorporates the
Dzyaloshinskii-Moriya and XY anisotropies. Doping with nonmagnetic
impurities, such as Zn, is considered according to classical
percolation theory, whereas the effect of Sr, which introduces
charge carriers into the {\cuoo} planes, is described as a dipolar
frustration of the antiferromagnetic order. We calculate several
magnetic, thermodynamic, and spectral properties of the system,
such as the antiferromagnetic order parameter, $\sigma_0$, the
N\'eel temperature, $T_N$, the spin-stiffness, $\rho_s$, and the
anisotropy gaps, $\Delta_{DM}$ and $\Delta_{XY}$, as well as their
evolution with both Zn and Sr doping. We explain the nonmonotonic
and reentrant behavior experimentally observed for $T_N(x,z)$ by 
H\"{u}cker {\it et al.} in Phys. Rev. B {\bf 59}, R725 (1999), as 
resulting from the reduction, due to the nonmagnetic impurities, 
of the dipolar frustration induced by the charge carriers 
(order-from-disorder). Furthermore, we find a similar nonmonotonic 
and reentrant behavior for all the other observables studied. Most 
remarkably, our results show that while for $x\approx 2\%$ and 
$z=0$ the Dzyaloshinskii-Moriya gap $\Delta_{DM}=0$, for $z=15\%$ 
it is approximately $\Delta_{DM}\approx 7.5$ cm$^{-1}$. The later is 
larger than the lowest low-frequency cutoff for Raman spectroscopy 
($\sim 5$ cm$^{-1}$), and could thus be observed in one-magnon 
Raman scattering.

\end{abstract}

\pacs{  } \maketitle

\section{Introduction}
High temperature superconductivity is obtained by introducing
charge carriers (holes or electrons) into Cu-based Mott-Hubbard
insulators. At half filling such systems exhibit long range
antiferromagnetic (AF) order in the ground state, which is however
rapidly suppressed by the introduction of charge carriers. Consider,
for example, the case of {\lco}, the simplest of the parent compounds.
The replacement of La$^{3+}$ by Sr$^{2+}$ ions in {\lasco},
through which holes are doped into the {\cuoo} planes, causes the
destruction of the canted N\'eel order already at
$x\approx0.02$,\cite{Kastner} showing that the doped holes strongly
frustrate the underlying AF order within the {\cuoo} planes.
Conversely, doping of iso-valent nonmagnetic impurities into
{\lco} is known to have less dramatic effects. For example, by
replacing Cu$^{2+}$ for Zn$^{2+}$ in {\lczo},\cite{Greven} the AF
order is suppressed at much higher impurity concentration and the
monotonic and rather smooth decrease of the N\'eel temperature
$T_N$ with $z$ can be described, at least in the limit of low
dilution, within classical percolation theory.

The combined effect of adding holes {\it and} nonmagnetic
impurities into {\lasczo} was studied experimentally by H\"ucker
{\it et al.}\cite{Hucker} It has been found that, even though
each kind of impurity independently suppresses the AF order, an
enhancement of the antiferromagnetism can occur when these
impurities are combined. For example, in the presence of Zn,
$z>0$, the AF order survives for $x>0.02$.\cite{Hucker} More
interestingly, it was also found that for $x=0.017$, the N\'eel
temperature exhibits a nonmonotonic behavior as a function of Zn
doping. In fact, $T_N$ is first enhanced from $125$ K at $z=0$ to
$144$ K at $z=0.05$, and only then starts to be suppressed by the
dilution (see Fig. 4 of Ref.\ \onlinecite{Hucker}).

This remarkable reentrant effect was immediately addressed
theoretically by Korenblit {\it et al.},\cite{Korenblit} who first
suggested that Zn reduces the frustration induced by Sr, through
the dilution of ferromagnetic bonds. These authors considered an
{\it isotropic Heisenberg spin system} and neglected the
Dzyaloshinskii-Moriya (DM) and pseudo-dipolar (XY) anisotropies,
including only the much smaller interplanar superexchange
$J_\perp$. More recently, however, the same authors suggested that
$J_\perp$ has little or nothing to do with the suppression of the
AF order in {\lasczo} ($J_\perp$ was shown to be almost $x$ and
$z$ independent), which should be, instead, solely determined by
{\it intraplanar} correlations.\cite{KorenblitInterlayer} This is
in agreement with the recent findings of Juricic {\it et al.},
\cite{VladimirHeli} where it has been shown that the robustness of
the canted N\'eel state is determined by the DM gap. The collinear
long range AF order is destroyed when the renormalized DM gap
vanishes, at $x\approx\Delta_{DM}/J\sim 0.02$, and for higher
doping the magnetism becomes incommensurate.
\cite{VladimirHeli,Luscher-IC} The DM and XY anisotropies have 
also been shown to be behind the unusual magnetic susceptibility 
response in {\lasco},\cite{Lavrov,Marcello} for a rather wide 
range in doping and temperature. In addition they are also 
responsible for the appearance of a field-induced mode in the 
one-magnon Raman spectrum of {\lasco}.\cite{Gozar,Marcello-Lara} 
Thus it is clear that any realistic description of the reentrant 
magnetism in {\lasczo} must properly take into account such 
anisotropies.

In this paper we revisit the problem of the reduction of
frustration in {\lasczo}, first proposed by Korenblit {\it et
al.},\cite{Korenblit} within the framework of a long-wavelength
nonlinear sigma model (NLSM) that properly includes DM and XY
anisotropies, dilution and frustration. We show that indeed
dilution weakens the frustration by reducing the dipolar-magnon
coupling constant. The result is a nonmonotonic behavior not only
for $T_N$, but also for other observables like: order parameter,
weak-ferromagnetic moment, and anisotropy gaps. Moreover, since 
the destruction of the AF order is directly related to the vanishing 
of the DM gap,\cite{VladimirHeli} we find that the reentrant 
antiferromagnetism is accompanied by a reentrant behavior for
the DM gap itself. Finally, we show that the NLSM formulation, 
together with classical percolation theory, can also describe 
the experiments in the highly Zn-diluted limit, provided the 
appropriate bond percolation factor is used.

The paper is organized as follows. In section \ref{Impurities} we
discuss the different impurities to be considered, we review the
derivation of the anisotropic NLSM for {\lco}, introduce dilution
by Zn according to the classical percolation theory, and include
the dipolar frustration within the framework of the Shraiman and
Siggia model.\cite{Shraiman} In section \ref{Framework} we present
our general model and compute all the renormalizations of the
physical quantities. In section \ref{Observables} we discuss the 
effect of the reduction of frustration in different physical 
quantities and propose experiments that could lead to the 
observation of these effects. This section is written in such a 
way that the reader not interested in the theoretical details can 
skip sections \ref{Impurities} and \ref{Framework}. In section 
\ref{Conclusions} we present our conclusions. In Appendix 
\ref{Appendix-A} we review the continuum limit for the site and 
bond percolation factors, while in Appendix \ref{Appendix-B} we 
provide a detailed derivation of the NLSM with dilution.

\section{Impurities in host $\mbox{\lco}$}

\label{Impurities}

\begin{figure}[t]
\includegraphics[scale=0.35,]{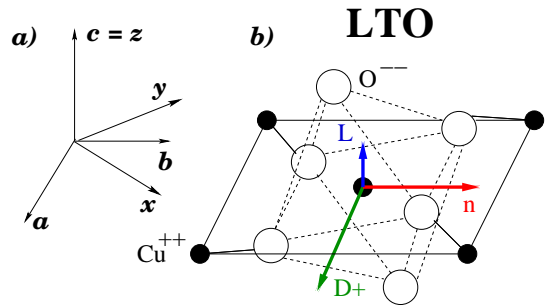}
\caption{Left: Tetragonal ($xyz$) and orthorhombic ($abc$)
coordinate systems. Right: In the LTO phase the tilting axis of
the CuO$_6$ octahedra, represented by the vector ${\bf D}_+$ (in
green), determines the orthorhombic $a$ axis. The DM and XY
anisotropies then determine that the staggered order parameter
${\bf n}$ (in red) is oriented along the orthorhombic $b$
direction while the weak-ferromagnetic moment ${\bf L}$ (in blue)
is perpendicular to the {\cuoo} plane.}
\label{Fig-Tilting}
\end{figure}

{\lco} is a layered antiferromagnet with a rather large XY
anisotropy which below $530$ K is in the low-temperature
orthorhombic (LTO) phase. Although the orthorhombicity resulting
from the staggered tilting of the CuO$_{6}$ octahedra is small, it
is responsible for the presence of an antisymmetric exchange of
the DM type once the spin-orbit coupling is considered. Together,
these two anisotropies, DM and XY, gap the transverse spin-wave
excitations along the $a$ and $c$ orthorhombic axis respectively, 
and lead to a canted antiferromagnetic ordering along the $b$ 
orthorhombic direction, see Fig.\ \ref{Fig-Tilting}b).

Theoretical studies of {\lco} concentrate on the Cu lattice,
because all the other atoms (La$^{3+}$, O$^{2-}$) are in a closed
shell configuration. In the crystal Cu$^{2+}$ has a missing
electron in the $d_{x^2-y^2}$ orbital, which carries a spin 1/2
and orders antiferromagnetically below $T_{N}$=325 K. There are
two kinds of impurities which, at some value of doping, destroy
the antiferromagnetic order in the layered La$_{2}$CuO$_{4}$
compound. Replacing La$^{3+}$ with Sr$^{2+}$ leads to a rapid
suppression of the antiferromagnetic order in the plane. This
happens because the holes donated to the planes form Zhang-Rice
singlets with the local moments of Cu$^{2+}$,\cite{ZhangRice} which
act as effective mobile holes in the spin lattice. Doping the 
crystal with Zn, i.e. substituting Cu$^{2+}$ with iso-valent 
Zn$^{2+}$, introduces {\it vacancies} at the Cu positions. In 
fact, Cu$^{2+}$ has an electronic configuration [Ar]3d$^{9}$, 
while Zn$^{2+}$ is [Ar]3d$^{10}$. This means that there is no 
magnetic moment at the Zn position and therefore the effect 
of Zn-doping is simply to dilute the spins within the plane.

The effect of the competition between Sr and Zn impurities in
La$_{2-x}$Sr$_{x}$Cu$_{1-z}$Zn$_{z}$O$_{4}$ was experimentally
investigated in Ref.\ \onlinecite{Hucker} and addressed
theoretically in Refs.\ \onlinecite{Korenblit} and \onlinecite{Vojta} . 
Both theoretical works have, as starting point, an isotropic 
quantum Heisenberg Hamiltonian on a square lattice. However, 
recent experimental\cite{Lavrov,Gozar,Reehuis} and theoretical
\cite{Papanicolaou,Marcello,Marcello-Lara,Gooding,Luscher-Neel} studies have 
emphasized the very important role of DM and XY anisotropies and
thus we shall now describe the effect of dilution and frustration 
in an effective model including anisotropies.

\subsection{Undoped {\lco}}

At long wavelengths, the isotropic Heisenberg Hamiltonian is well
described, in the paramagnetic phase, by a $O(3)$
NLSM.\cite{Chakravarty} In this paper, however, we use a
generalized NLSM, derived by Chovan and Papanicolaou in 
Ref.\ \onlinecite{Papanicolaou} and Silva Neto {\it et al}. 
in Ref.\ \onlinecite{Marcello}, which includes the DM and XY 
anisotropies. The action of the model reads
\begin{eqnarray}
S_{tot} &=& \int d \tau
[L_{H}+L_{DM}+L_{XY}+L_{WZ}], \nonumber \\
&=& S^{2} \sum_{<i,j>}
J{\bf \Omega}_{i}{\bf \Omega}_{j} + {\bf D}_{ij} \cdot ({\bf
\Omega}_{i} \times{\bf \Omega}_{j})
+{\bf \Omega}_{i}\widehat{\Gamma}_{ij}{\bf \Omega}_{j} \nonumber \\
&-& iS \sum_{j \epsilon 2Dlattice} \delta{\bf
\Omega}_{j}\cdot ( {\bf \Omega}_{j} \times
\partial_{0}{\bf \Omega}_{j}).
\label{Action}
\end{eqnarray}
In Eq.\ (\ref{Action}), $L_{H}$, $L_{DM}$, $L_{XY}$ and
$L_{WZ}$ are, respectively,
the Heisenberg, DM, XY, and Wess-Zumino terms, ${\bf \Omega}$ is a
unit vector along the spin direction, ${\bf S}=S{\bf \Omega}$, and
$S=1/2$. The super-exchange parameter is denoted by $J$. The DM
vector ${\bf D}_{ij}$ and the anisotropy matrix
$\widehat{\Gamma}_{ij}$ are defined on the Cu-Cu
bonds.\cite{Shektman,Koshibae} In the $(x,y)$ coordinate system, 
see Fig.\ \ref{Fig-Tilting},
${\bf D}_{x}=(0,d,0)$, ${\bf D}_{y}=(d,0,0)$,
$\widehat{\Gamma}_{x}=diag(\Gamma_{1}+\Gamma_{2},
\Gamma_{1}-\Gamma_{2},\Gamma_{3})$, and
$\widehat{\Gamma}_{y}=diag(\Gamma_{1}-\Gamma_{2},
\Gamma_{1}+\Gamma_{2},\Gamma_{3})$.
Here, the index $x$ means along the horizontal Cu-O-Cu bonds, $y$ 
means along the vertical Cu-O-Cu bonds, $d\sim10^{-2}J$, 
$\Gamma_{1,2,3}\sim10^{-4}J$, and $\Gamma_{1}>\Gamma_{3}$.

Our next step is to separate the fast ${\bf l}$ and slow ${\bf n}$
varying spin components
\begin{eqnarray}
{\bf \Omega}_{i}&=&
(-1)^{i}{\bf n}_{i}\sqrt{1-(a{\bf l}_{i})^{2}}+a{\bf l}_{i} \nonumber \\
& \simeq & (-1)^{i}{\bf n}_{i}+a{\bf l}_{i}-
\frac{(-1)^{i}}{2}a^{2}{\bf n}_{i}{\bf l}_{i}^{2}.
\label{Decomposition}
\end{eqnarray}
In Eq.\ (\ref{Decomposition}) $a$ is the lattice constant and $i$
is an index on the 2D lattice, $i=(p,q)$; $(-1)^{i}$ should be
understood as $(-1)^{p+q}$, ${\bf n}_{i}$ as ${\bf n}_{p,q}$.
After substituting Eq.\ (\ref{Decomposition}) into Eq.\
(\ref{Action}) we find
\begin{eqnarray}
\nonumber S_{tot} &=&  \int d^{2}{\bf r} \Big\{
\frac{JS^{2}}{2}[(\nabla {\bf n})^{2}+8{\bf l}^{2}]+
\frac{4S^{2}}{a}{\bf d_{+}}\cdot({\bf n}\times{\bf l}) \nonumber \\
&+&\frac{2S^{2}}{a^{2}} (\Gamma_{1}-\Gamma_{3})n_{z}^{2}-
\frac{iS}{a} {\bf l}\cdot({\bf n}\times{\dot{{\bf n}}})
\Big\},
\label{Stot}
\end{eqnarray}
where ${\bf d}_{+}=({\bf D}_{x}+{\bf D}_{y})/2$.

\subsection{Effect of non-magnetic impurities - Zn doping}

Recently, an effective field theory for the Heisenberg
antiferromagnet with non-magnetic impurities was derived by Chen
and Castro Neto.\cite{ChenPaper} While in Ref.\ \onlinecite{ChenPaper}
the authors only considered the {\it isotropic} case, here we 
generalize the results of Ref.\ \onlinecite{ChenPaper} to include 
the DM and XY anisotropies.

We introduce a function $p_{i}$, which is 1 on a Cu site and zero
on a Zn site, with the property $p_{i}^{2}=p_{i}$. We then assume
this function to be smooth and expand $p_{j}$ in the neighborhood
of its nearest neighbor site. In a continuum limit, $p_{i}$ and
$K_{ij}=p_{i}p_{j}$ are replaced, respectively, by $p({\bf r})$
and $K({\bf r})$, which are the so called site and bond
percolation factors. As it is shown in the Appendix
\ref{Appendix-A}, for the case of a homogeneous distribution of
static impurities, they can be replaced by their average values,
$P_{\infty}(z)=1-z$ and $K(z)=1-3z$. In the following we omit the
$z$-dependence in $P_{\infty}$ and $K$ for the sake of shortening
the notation.

In the presence of dilution, the action (\ref{Action}) will be
modified as follows: the terms $L_{H}, L_{DM}$, and $L_{XY}$,
which contain $\sum_{<i,j>}f({\bf \Omega}_{i}, {\bf \Omega}_{j})$,
will be multiplied by $K$. The Wess-Zumino term, on the other
hand, contains ${\bf \Omega}^{3}_{i}$ and therefore it is
proportional to $p^{3}({\bf r})=p({\bf r})$. Thus, within the
homogeneous distribution assumption, it will be multiplied by
$P_{\infty}$. In this way, $K$ and $P_{\infty}$ appear simply as
prefactors of the relevant integrals.\cite{ChenPaper} A detailed
derivation of the NLSM in the presence of nonmagnetic impurities
is given in Appendix \ref{Appendix-B}.

The next step is to integrate out ${\bf l}$ in Eq.\ (\ref{Stot})
to obtain the action in the presence of dilution (see Appendix 
\ref{Appendix-B})
\begin{widetext}
\beq
S_{Zn}=
\frac{1}{2gc(K/P_\infty)} \int \rmd \tau \int \rmd^{2}{\bf r} \left\{
(\partial_{\tau} {\bf n})^{2}
+Z[c^{2}(\nabla {\bf n})^{2}+m_{a}^{2}n_{a}^{2}+m_{c}^{2}n_{c}^{2}]
\right\},
\label{Action-Diluted}
\eeq
\end{widetext}
where we have defined $Z=K^2/P_\infty$, $g$ is the usual NLSM 
coupling constant, and $c$ is the spin-wave velocity. 
Here we used that $\sqrt{2gcS^{2}/Ja^{2}}{\bf d}_{+} = 2 \sqrt{2} 
Sd \overrightarrow{e_{a}} = m_{a}\overrightarrow{e_{a}}$ and
$(4gcS^{2}/a^{2})(\Gamma_{1}-\Gamma_{3})=32JS^{2}(\Gamma_{1}-
\Gamma_{3})=m_{c}^{2}$, where $\overrightarrow{e_{a}}$ is the unit 
vector along the $a$ orthorhombic direction. Therefore, the last two
terms in Eq.\ (\ref{Action-Diluted}) correspond, respectively, to
the DM and XY anisotropy gaps, showing that the spin ordering has 
an easy-axis along the orthorhombic $b$ direction, see Fig.\
\ref{Fig-Tilting}.

From the above action it is clear that when only Zn impurities
are doped into {\lczo} the two anisotropy gaps renormalize according 
to
\beq
M_{a,c}=\sqrt{Z}\;m_{a,c},
\eeq
and thus decrease rather smoothly with dilution and vanish at the 
percolation threshold.

In what follows we shall switch freely between $\Delta_{DM}$ and
$M_a$, when refering to the DM or in-plane gap, and also between 
$\Delta_{XY}$ and $M_c$, when refering to the XY or out-of-plane
gap, withouth any loss of generality.

\subsection{Derivation of the N\'{e}el temperature}

Let us now derive an expression for the N\'eel temperature
in terms of the anisotropy gaps and spin-stiffness, renormalized 
by dilution. Starting from the action in 
Eq.\ (\ref{Action-Diluted}), we split the {\bf n}-field into 
its longitudional $\sigma_{0}$ and transverse {${\bf n_{\perp}}$ 
components, ${\bf n}=(n_{a},\sigma_{0},n_{c})$. Here 
${\bf n_{\perp}}=(n_{a},n_{c})$, and $\sigma_{0}=const.$ is 
the order parameter. The action then reads
\begin{equation}
S^{d}[{\bf n_{\perp}}]=\frac{1}{2gc}
\frac{1}{\beta}\sum_{\omega_{n}}\int \frac{d^{2}{\bf k}}{(2
\pi)^{2}} {\bf n_{\perp}} \widehat{A}^{d} {\bf n_{\perp}},
\end{equation}
with
\begin{equation}
\widehat{A}^{d}=\textbf{I}_{2}\Big[(P_{\infty}/K)
\omega_{n}^{2}+Kc^{2} \textbf{k}^{2} \Big]+
\mbox{diag}[Km_{a}^{2},Km_{c}^{2}],
\end{equation}
where $\omega_n=2\pi n/\beta$ are the Matsubara frequencies, 
${\bf I}_{2}$ is a 2D identity matrix, and $\beta=1/k_BT$
is the inverse temperature. Defining Tr as $\beta^{-1}
\sum_{\omega_{n}}\int d^{2}{\bf k}/(2 \pi)^{2}$ and introducing
the fixed length constraint into the action through a Lagrange
multiplier, $\lambda$, the partition function can be written as
\begin{eqnarray}
Z &=& \int D {\bf n} \mbox{ } \delta ({\bf n}^{2}-P_{\infty})
\exp \{-S^{d}[{\bf n}] \} \nonumber \\
&=& \int D \sigma_{0} D
{\bf n_{\perp}} D \lambda \exp \Big\{-\frac{1}{2gc}
\mbox{Tr}[{\bf n_{\perp}} \widehat{A}^{d} 
{\bf n_{\perp}} \nonumber \\
&+& i \lambda (\sigma_{0}^{2}+
{\bf n_{\perp}}^{2}-P_{\infty})] \Big\} \nonumber \\
&=& \int D \sigma_{0} D \lambda
\exp\{-S_{eff}[\lambda,\sigma_{0}]\},
\label{PartFunc}
\end{eqnarray}
where the effective action reads
\begin{equation}
S_{eff}[\lambda, \sigma_{0}]=\frac{1}{2gc} \mbox{Tr}[i \lambda
\sigma_{0}^2-i \lambda P_{\infty}] +\frac{1}{2} \sum_{\alpha=a,c}
\mbox{Tr} \ln [A_{\alpha \alpha}^{d}+i \lambda].
\label{EffAction}
\end{equation}
In Eqs.\ (\ref{PartFunc}) and (\ref{EffAction}) we used that the
average length of spin per lattice site is $P_{\infty}$ in the
diluted case.\cite{ChenPaper,ChenThesis}

The thermodynamic properties of the diluted system can be
determined, at the mean field level, by solving the saddle point
equations of the effective action (\ref{EffAction}). From $\delta
S_{eff}[\lambda,\sigma_{0}]/ \delta \lambda=0$ we find
\begin{equation}
\frac{P_{\infty}-\sigma_{0}^{2}}{gc}=\sum_{\alpha=a,c}\mbox{Tr}
\frac{1}{A_{\alpha \alpha}^{d}+i \lambda}.\label{Saddle-Point}
\end{equation}
The only
condition for the trace to converge is to choose
$i\lambda=m_{0}^{2}$, a real positive number. As usual, we
interpret it as the inverse correlation length, $m_{0}=1/\xi$.

At the N\'{e}el critical point the order parameter $\sigma_{0}$ 
vanishes, the correlation length $\xi$ diverges, and the 
equation for the critical temperature is
$$
\frac{\beta P_{\infty}^{2}}{gcK}=\sum_{\alpha=a,c}
\sum_{\omega_{n}} \int \frac{d^{2}{\bf k}}{(2 \pi)^{2}}
\frac{1}{\omega_{n}^{2}+Z c^{2}({\bf k}^{2}+m_{\alpha}^{2}/c^{2})}.
$$
After performing the Matsubara summation and the integration over
momenta, the last expression simplifies to
\begin{equation}
4\pi\beta_N P_{\infty}K \rho_{s}+\sum_{\alpha=a,c}\ln \left[ 2 \sinh
\Big(\frac{\beta_N}{2}
\sqrt{Z}\;m_{\alpha}\Big)
\right]=0,\label{T-Neel}
\end{equation}
where $\beta_N=1/k_B T_N$ and we defined the renormalized spin 
stiffness for the diluted system
\beq
\rho_{s}=c\left(\frac{1}{2g}-\frac{1}{g_{c}KP_{\infty}}\right),\eeq
with $g_{c}$ defined through the ultra-violet momentum cut-off
$\Lambda=4\pi/g_{c}$, as usual.

%
\begin{figure}[h]
\includegraphics[scale=0.4,angle=-90]{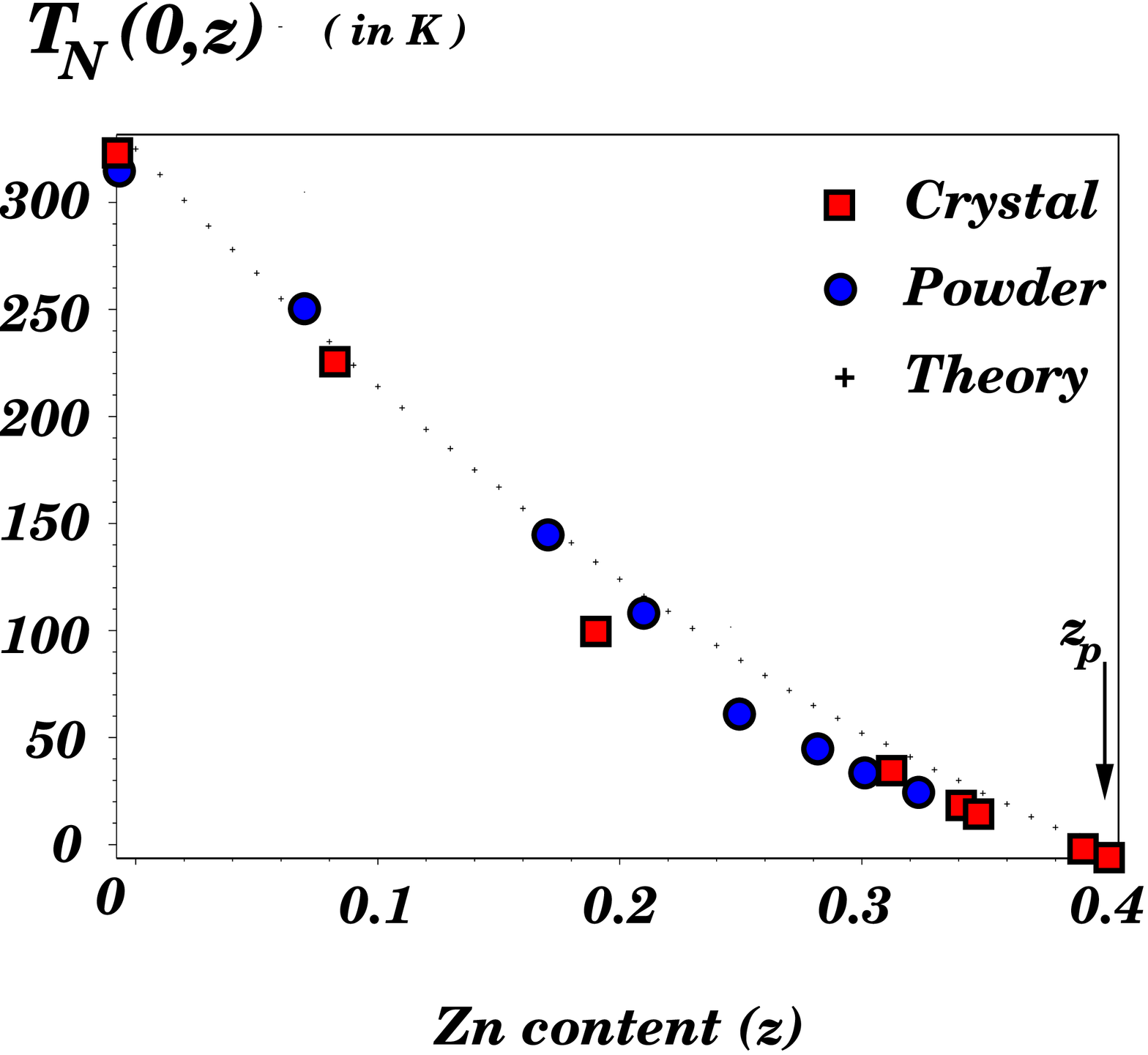}
\caption{N\'eel temperature (in K) as a function of Zn concentration
for $x=0$ up to the highly diluted regime. The experimental data
for either powder and crystal samples was taken from Refs.\ 
\onlinecite{Greven} and \onlinecite{Hucker}.}
\label{greven}
\end{figure}
%

The bond percolation factor $K(z)=1-3z$, proposed in Ref.\
\onlinecite{ChenPaper}, (see Appendix \ref{Appendix-A}) describes
well the experimental data for $T_{N}(z)$ only in the low doping 
range.\cite{Hucker,ChenPaper} For highly diluted samples, however, 
single crystals were not available before the work reported in 
Ref.\ \onlinecite{Greven}. The experiments were usually performed 
with single crystal samples below the doping threshold of 
$z\approx20\%$, while above 25$\%$ powder samples were used. 
After Ref.\ \onlinecite{Greven}, which used both single crystals
and powder samples, our conclusion is that for the heavily doped 
samples the bond percolation factor should be modified according 
to Refs.\ \onlinecite{BrooksHarris} and \onlinecite{WL}, which 
suggest that $K(z)=1-\pi z+\pi z^{2}/2$. In this Watson-Leath 
(WL) prescription, $T_{N}$ is zero at the percolation threshold, 
$z_p$. Theoretical, experimental, and numerical studies of highly 
diluted La$_{2}$Cu$_{1-z}$[Mn,Zn]$_{z}$O$_{4}$ are presented in 
Ref.\ \onlinecite{Greven}. There the authors describe an experiment
where the critical point is found at $z=42\%$. As it is observed 
in Fig.\ \ref{greven}, using the bond percolation factor as in the WL 
prescription, we obtain the correct description in the heavily 
doped regime. 

For the plot in Fig.\ \ref{greven} we used $J=100$ 
meV, which is much smaller than the real value of $\approx 135$ 
meV, but is the value that gives $T_N(x=0,z=0)=325$ K within our
saddle-point approximation. The inclusion of fluctuations away
from the saddle-point can allow for the more realistic value of 
$J$ to be used. In the next section we discuss the effect of Sr 
doping and in the following we shall concentrate our studies on 
single crystals doped simultaneously with Sr and Zn. 

\subsection{Effect of frustration - Sr doping}

In order to incorporate the effect of Sr doping in
La$_{2}$Cu$_{1-z}$Zn$_{z}$O$_{4}$, we adopt the Shraiman and
Siggia model.\cite{Shraiman} In this model the holes introduced in
the system (La$_{2-x}$Sr$_{x}$Cu$_{1-z}$Zn$_{z}$O$_{4}$) via Sr
doping are represented by an effective dipolar field that couples
to the background magnetization current.\cite{Shraiman} In Ref.\
\onlinecite{VladimirHeli} it was shown that this interaction leads
to a reduction of the magnon gaps and spin stiffness, in agreement
with the experiments.\cite{Gozar} As we shall now demonstrate,
such reduction is not as strong if the compound is additionally
doped with Zn, since in the presence of non-magnetic impurities
the effective dipole-magnetization-current interaction should be
multiplied by the bond dilution factor. This mechanism of
reduction of frustration by nonmagnetic impurities has been
considered earlier by Korenblit {\it et al.} \cite{Korenblit}
within the isotropic O(3) NLSM. Here we discuss the role of
anisotropies.

In the Shraiman-Siggia model the interaction between magnons and
the dipolar field {\it in the absence of dilution} can be written 
as
\beq S_{int}=-2\lambda\int\rmd\tau\int\rmd^{2}{\bf r}
\;{\bf P}_{\mu}\cdot{\bf n}\times\partial_{\mu}{\bf n},
\label{Dip-Mag}
\eeq
where
\beq {\bf P}_{\mu}=i\partial_{\mu}\overline{\Psi}
\overrightarrow{\sigma}\Psi+\mbox{H.c.} ,\eeq
${\bf P}_{\mu}$ is the dipolar field representing the spin
current of the holes, $\Psi$ is the spinor wave function of the 
doped holes with dispersion centered at $(\pi/2,\pm \pi/2)$ and 
symmetry related points in the Brillouin zone, 
$\overrightarrow{\sigma}$ are the three Pauli matrices, and 
$\mu$ is a lattice index.

In the presence of dilution the above dipole-magnon interaction
will also have to be modified following the procedure described in
Appendix \ref{Appendix-B}. Essentially, one can represent the
semiclassical background spin distortion as a slowly varying SU(2)
rotation of the N\'eel state ${\bf R_{r}}$. The hopping term for
the doped holes involves the product 
${\bf R}_{r}{\bf R}_{r+{\bf a}}^{+}$, and since these are rotations 
in neighbouring sites, i.e. ${\bf R}_{i}{\bf R}_{j}^{+}$, in the 
presence of dilution this product has to be replaced by
\beq 
{\bf R}_{i}{\bf R}_{j}^{+}\rightarrow
{\bf R}_{i}{\bf R}_{j}^{+}p_{i}p_{j}, 
\label{SU2-Rot}
\eeq
where $i,j$ are nearest neighbor sites. Consequently, the coupling 
constant $\lambda$ between the dipolar field and the background 
magnetization current in Eq.\ (\ref{Dip-Mag}) should be changed
according to 
\beq
\lambda \rightarrow K\lambda, 
\label{Klambda}
\eeq
because (\ref{Dip-Mag}) comes from (\ref{SU2-Rot}).\cite{Shraiman}
Here $K$ is the bond dilution factor which in the homogeneous
approximation is given by $K=<p_{i}p_{j}>$.

%
\begin{figure}[t]
\includegraphics[scale=0.3]{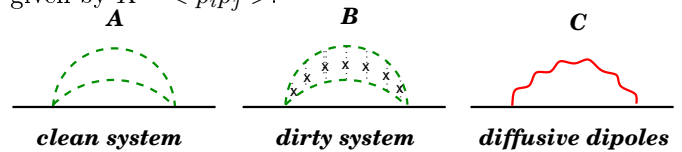}
\caption{Contribution to the magnon propagator from the
dipolar fields. The figure on the left (A) contains the
polarization diagram of the doped holes. In (B) we introduce
disorder through the scattering of the holes by impurities 
and this is represented effectively by the magnon-dipole
bubble diagram (C) where we use the diffusive dipole propagator.}
\label{lindhard}
\end{figure}
%

\section{General model}

\label{Framework}

The complete action describing the simultaneous effect of 
magnetic dilution by Zn and frustration by Sr reads
\begin{widetext}
\beq
S=
\frac{1}{2gc(K/P_\infty)} \int \rmd \tau \int \rmd^{2}{\bf r} \left\{
(\partial_{\tau} {\bf n})^{2}
+Z[c^{2}(\nabla {\bf n})^{2}+m_{a}^{2}n_{a}^{2}+m_{c}^{2}n_{c}^{2}]
\right\}-2\lambda K\int\rmd\tau\int\rmd^{2}{\bf r}
\;{\bf P}_{\mu}\cdot{\bf n}\times\partial_{\mu}{\bf n}+S_{d},
\eeq
\end{widetext}
where 
\beq
S_d=\frac{1}{2}\int\rmd\tau\int\rmd^{2}{\bf r}\;
{\bf P}_{\mu} \, G_D^{-1} \, {\bf P}_{\mu}
\eeq
describes the fluctuations of the dipolar field. Since
the dipolar field is a pseudo-fermionic composite field, it is an
operator that represents the spin current of the doped holes,
the dipolar susceptibility (inverse dipole propagator) can be 
obtained from a polarization diagram (see Fig.\ \ref{lindhard}). 

In what follows we adopt the same procedure used by Sachdev in 
Ref.\ \onlinecite{Sachdev} and we use the results from the Fermi 
liquid polarization diagram in two dimensions. However, while 
in Ref.\ \onlinecite{Sachdev} the ballistic limit (clean system)
was used in the calculation of the polarization diagram, see
left diagram in Fig.\ \ref{lindhard}, here we consider the 
diffusive limit (dirty system) because of the scattering of the 
doped holes by the Sr impurities, see middle diagram in 
Fig.\ \ref{lindhard}. This assumption is consistent with
recent results from L\"uscher {\it et al.}\cite{Luscher-Neel} 
who have shown that the dynamics of these dipolar fields is 
highly diffusive. Thus, we shall write for the dipole propagator 
in the diffusive limit (see right diagram in Fig.\ \ref{lindhard})
\beq G_D({\bf q},\omega_n)=\kappa_{d} \frac{D{\bf q}^2 }{D{\bf
q}^2+|\omega_n|},
\label{Prop2}
\eeq
where $\kappa_d$ is proportional to the inverse static 
susceptibility and $D$ is the diffusion constant, which
is assumed to be large.

Because of the peculiar dipolar-magnon coupling from 
Eq.\ (\ref{Dip-Mag}), the magnon propagator will be
renormalized by the fluctuations of the dipolar field
(see Fig.\ \ref{lindhard}).
The self-energy correction to the magnon propagator is
\begin{widetext}
\beq \Sigma_{M}^{\alpha}(\qb,\on)=\kappa_d (\lambda^{2}K^{2})
\frac{gcK}{\pinf} \eabc^{2} \trace
\frac{(\kb+2\qb)^{2}}{\on^2+Zc^{2}(\kb+\qb)^{2}+Zm_{\beta}^{2}}
\frac{D{\bf k}^2 }{D{\bf k}^2+|\omega_n|}, 
\eeq
where $\eabc$ is the completely antisymmetric tensor.

By summing up the one-loop corrections to the magnon Green's
function, we obtain
\beq \widetilde{G}_{M} = G_{M}\sum_{i=0}^{\infty}(\Sigma_{M}
G_{M})^{i} = [G_{M}^{-1}-\Sigma_{M}]^{-1}. 
\eeq 
Writing this expression explicitly, we get for the magnon 
propagator 
\beq
\widetilde{G_{M}^{\alpha}}^{-1}(\qb,\on)=
(gcK/\pinf)^{-1}\Big(\on^2+Zc^{2}\qb^{2}+Zm_{\beta}^{2}\Big)
-\Sigma_{M}({\bf
0})-\frac{1}{2}q_{\mu}q_{\nu} \frac{\partial^{2} \Sigma({\bf
q})}{\partial q_{\mu} \partial_{\nu}} \Big|_{{\bf q}=0}, 
\eeq
where we expanded the self-energy around zero momentum up to the
second order term, in order to obtain a correction to the gaps and
spin stiffness, and thus
\beq
\widetilde{G_{M}^{\alpha}}^{-1}(\qb,\on)=\left(\frac{gcK}{\pinf}\right)^{-1}\on^2+
\left[ \frac{Kc}{g}-\frac{\delta_{\mu\nu}}{2}
\frac{\partial^{2}\Sigma_{M}^{\alpha}(\qb,0)}{\partial
q_{\mu}\partial q_{\nu}}\Big|_{\qb=0} \right]\qb^2
+\left(\frac{gcK}{\pinf}\right)^{-1} \left[
Zm_{\alpha}^{2}-\frac{gcK}{\pinf}\Sigma(0,0) \right].\eeq
From the above equation the expressions for the mass and the
spin stiffness renormalizations are readily derived
\begin{equation}
M_{\alpha}^{2}(x,z)=Zm_{\alpha}^{2}-\frac{gcK}{\pinf}\Sigma(0,0)\quad
\mbox{and}\quad
\rho_{s}(x,z)=K\rho_{s}-\frac{\delta_{\mu\nu}}{2}
\frac{\partial^{2}\Sigma_{M}^{\alpha}(\qb,0)}{\partial
q_{\mu}\partial q_{\nu}}\Big|_{\qb=0},
\end{equation}
where $c/g$ is the bare spin stiffness in a clean system. Explicitly,
the mass renormalization due to Sr and Zn impurities reads
\beq 
M_{\alpha}^{2}(x,z)= Z\left[ m_{\alpha}^{2}-
\kappa_d(gc\lambda)^2 \eabc^{2} Z I(z) \right],
\eeq
where
\beq
I(z)=\trace \frac{\kb^2}{\on^2+Zc^2 \kb^2
+Zm_{\beta}^{2}}\frac{D{\bf k}^2 }{D{\bf
k}^2+|\omega_n|}. 
\eeq
The spin stiffness, on the other hand, renormalizes according 
to the formula
\bea \mbox{ } \\ \nonumber \rho_{s}(x,z)=K\rho_{s}\left[ 1
-\frac{1}{2\beta}\sum_{\omega_n}\int \frac{d^2 \kb}{(2\pi)^2}
\left( \frac{\partial^2}{\partial
\qb^2}\frac{(\kb+2\qb)^2}{\on^2+Zc^2(\kb+\qb)^2+Zm_{\beta}^{2}}\Big|_{\qb=0}
\right)\frac{D{\bf k}^2 }{D{\bf k}^2+|\omega_n|} \right]. \eea
The new momentum cut-off $\Lambda$ for the dipoles, which
renormalizes the spin stiffness, is set by $k_{F}=\sqrt{\pi x}$,
because our theory should be valid at distances much larger than
the average distance between the Sr impurities.

In the zero temperature, $T\rightarrow 0$, and highly diffusive,
$D\rightarrow\infty$, limits the expressions for $M_\alpha$ and $\rho_s$ read
\bea M_{\alpha}^{2}[x,z;T\rightarrow 0]&=&Z\left\{
m_{\alpha}^2-\kappa_d(gc\lambda)^{2}\eabc^2 Z^{1/2} \frac{m_{\beta}^3}{4\pi c^4}
\left[ \frac{1}{3}\Big[ \Big( 1+\kcm \Big)^{3/2}-1 \Big]- \Big[
\Big( 1+\kcm \Big)^{1/2}-1 \Big] \right]  \right\}, 
\label{Mass-Ren}
\eea
and
\beq \rho_{s}[x,z;T\rightarrow0]=K\rho_{s}\left\{ 1-
\frac{\kappa_d(gc\lambda)^{2}\eabc^2 Z^{1/2}m_{\beta}}{4\pi c^4}\left[ 1+\frac{\Big(
\kcm \Big)^2+\frac{1}{2}\kcm-1}{\Big( 1+ \kcm \Big)^{3/2}} \right]
\right\}. 
\label{Stiff-Ren}
\eeq
\end{widetext}
%


\section{Observables}

\label{Observables}

Eqs. (\ref{Mass-Ren}) and (\ref{Stiff-Ren}) are the most important
results of this paper. As we can clearly see, although both $M_\alpha$
and $\rho_s$ are reduced when only Sr ($\lambda\neq 0$ and $Z=1$) or 
only Zn ($\lambda=0$ and $Z<1$) are doped {\it independently} into 
{\lco}, when combined a nonmonotonic and reentrant behaviour can 
indeed occur due to the $\lambda^2\sqrt{Z}$ coeficient in the 
self-energy correction (see Eqs. (\ref{Mass-Ren}) and (\ref{Stiff-Ren})).

In what follows we discuss the evolution with both Zn and Sr 
doping of several spectral, thermodynamic, and magnetic 
properties of the {\lasczo} system using the results obtained 
from the previous section. As we shall see, the nonmonotonic 
behaviour of different physical observables, such as the order 
parameter, the N\'eel temperature, and the weak-ferromagnetic 
moment, all follow from the nonmonotonicity induced by the 
competition between dilution and frustration in $M_\alpha$
and $\rho_s$.

\subsection{Dzyaloshinskii-Moriya or in-plane gap}

Juricic {\it et al.}\cite{VladimirHeli} have recently studied the
evolution of the DM gap with Sr in {\lasco} (see also L\"uscher {\it
et al.}\cite{Luscher-Neel}). One of the most important findings
reported in Ref.\ \onlinecite{VladimirHeli} was that the DM gap
gives {\it robustness} to the canted N\'eel state and vanishes
at a critical concentration given by $x=const.\;(\Delta_{DM}/J)\approx 2\%$,
where $const.$ is $O(1)$ (see also curves in Figs. \ref{DM-Zn-fixed}
and \ref{DM-Sr-fixed}). The mechanism for the reduction of the 
DM gap was traced back to the self-energy corrections due to the 
dipolar frustration introduced by Sr doping, $M_\alpha$ with $Z=1$
($z=0$), and the theoretical curved was shown to agree quite well 
with the one-magnon Raman experiments of Gozar {\it et al.}\cite{Gozar}

%
\begin{figure}[htb]
\includegraphics[scale=0.35,angle=-90]{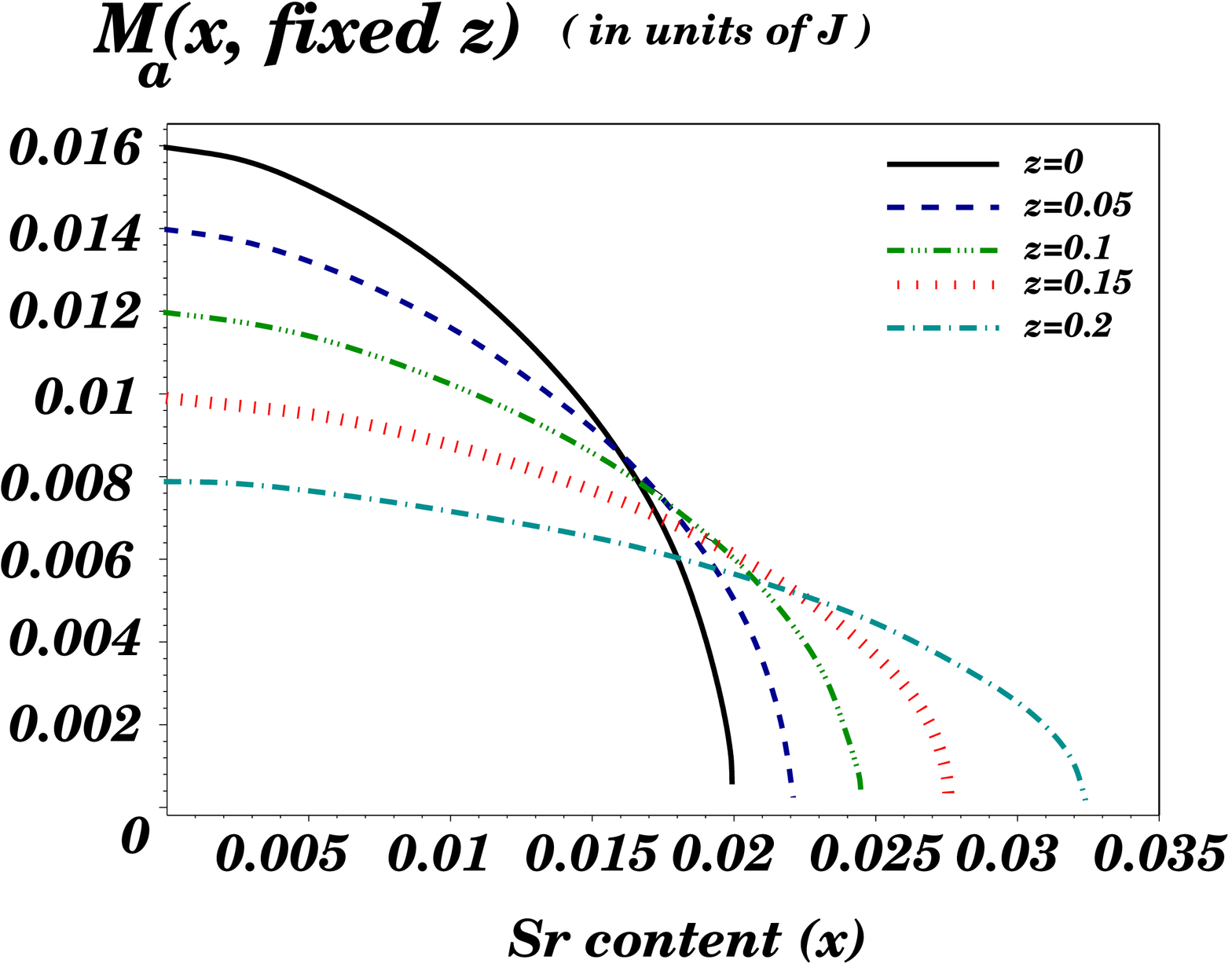}
\caption{Dependence of the DM gap ($\Delta_{DM}=M_a$, in units
of $J$) on Sr doping $x$ at various fixed Zn concentration: 
$z=0,0.05,0.1,0.15,0.2$. Although Zn and Sr independently 
reduce $M_a$, when combined an enhancement of $M_a$ does
occur.}
\label{DM-Zn-fixed}
\end{figure}
%

In fact, these experiments, performed at $10$ K, show that at 
$x=z=0$ the Dzyaloshinskii-Moriya (or in-plane) gap is $m_a=17.5$ 
cm$^{-1}$ (or $2.16$ meV), whereas it is reduced by almost $30\%$ 
at $x=1\%$, and it vanishes at $x=2\%$. In Fig.\ \ref{DM-Zn-fixed} 
we exhibit the Sr dependence of the DM gap for various fixed Zn 
concentrations (the DM gap is given in units of $J$). As it is 
evident from the plot, for small $x$ and $z$ the DM gap is always 
reduced. At large $x$, however, a reentrant behaviour is observed
already for very small Zn concentration. This effect is largest 
close to the Zn-free critical concentration $x=2\%$. In fact, 
while for $x=2\%$ and $z=0$ the DM gap is zero\cite{VladimirHeli,Gozar}, 
we find that for $x=2\%$ and $z=15\%$ the DM gap is 
$\Delta_{DM}\approx 7.5$ cm$^{-1}$, right above the lowest 
low-energy cutoff for one-magnon Raman spectroscopy 
($\sim 5$ cm$^{-1}$). Thus, we predict that low-energy Raman 
experiments in {\lasczo} samples with $x\approx 2\%$ and 
$z=15\%$ should observe a clear signal of a one-magnon mode 
with energy close to $7.5$ cm$^{-1}$ in the B$_{1g}$ scattering 
geometry,\cite{Marcello-Lara,LaraRaman} such as the ones performed 
by Gozar {\it et al.}\cite{Gozar} in Zn-free samples.

%
\begin{figure}[t]
\includegraphics[scale=0.35,angle=-90]{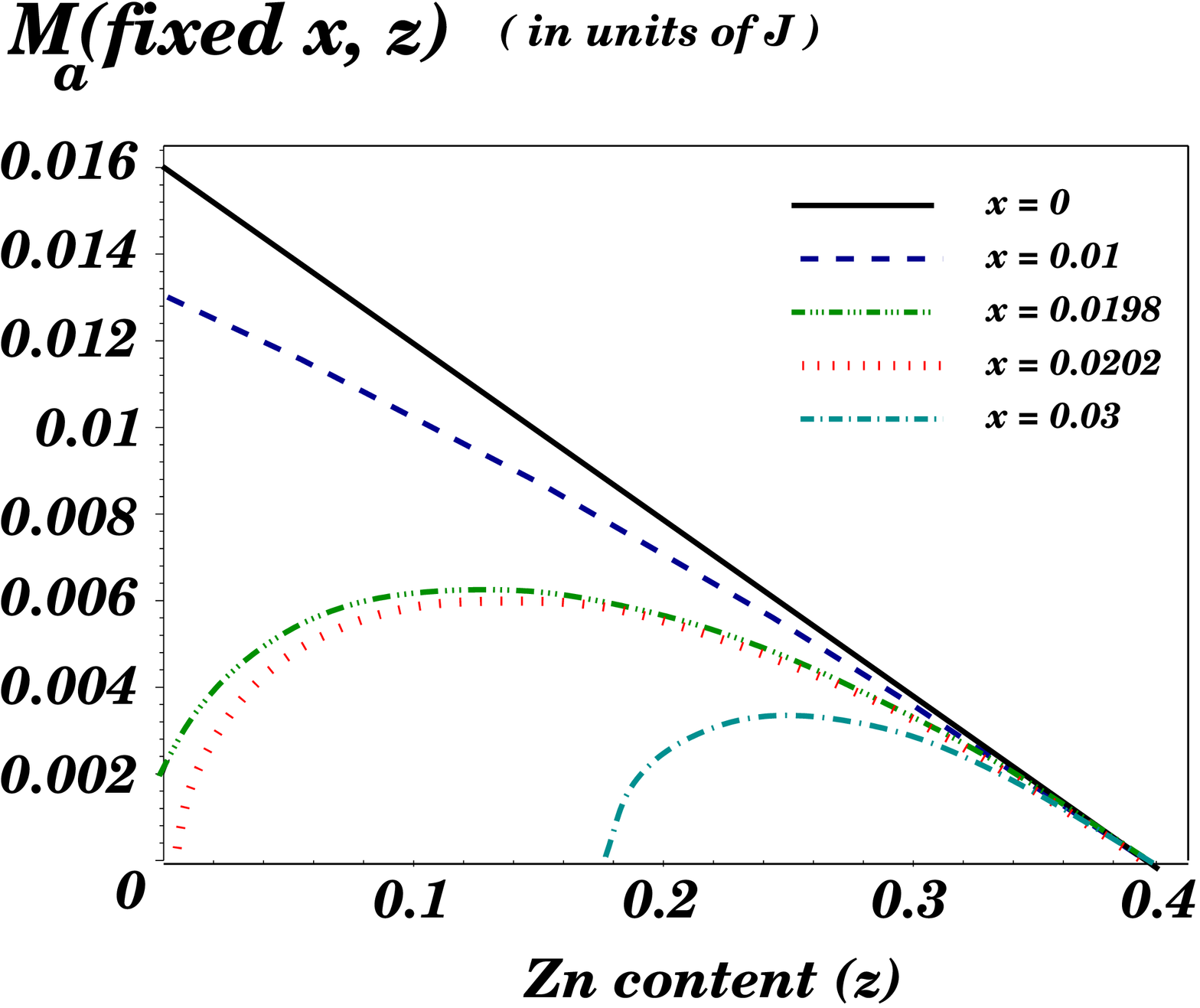}
\caption{Dependence of the DM gap ($\Delta_{DM}=M_a$, in units
of $J$) on Zn doping $z$ at various fixed Sr concentrations: 
$x=0,0.01,0.0198,0.0202,0.03$. The nonmonotonic and reentrant
behaviours for $M_a$ is evident.}
\label{DM-Sr-fixed}
\end{figure}
%

Fig. \ref{DM-Sr-fixed} shows the evolution of the DM gap as
a function of Zn doping for various fixed Sr concentrations. 
The nonmonotonic (for $x=0.0198$) and reentrant (for $x\geq 0.02$) 
behaviours predicted by our theory for the DM gap are also 
evident from this plot. As expected, all curves collapse into 
a single curve in the highly Zn-diluted regime. For completeness 
we show in Fig.\ \ref{DM-3D} a 3D plot with the evolution of 
the DM gap with both Zn and Sr doping.

As we have discussed in the previous section, the key mechanism 
for such nonmonotonic and reentrant behaviour observed in the DM
gap, through dilution by Zn, is the {\it decrease} of the effective 
dipolar-magnon coupling constant $\lambda\rightarrow K\lambda$ in 
Eq.\ (\ref{Klambda}), which therefore reduces the self-energy 
corrections to the magnon gaps, see (Eq.\ \ref{Mass-Ren}). 

%
\begin{figure}[b]
\includegraphics[scale=0.35,angle=-90]{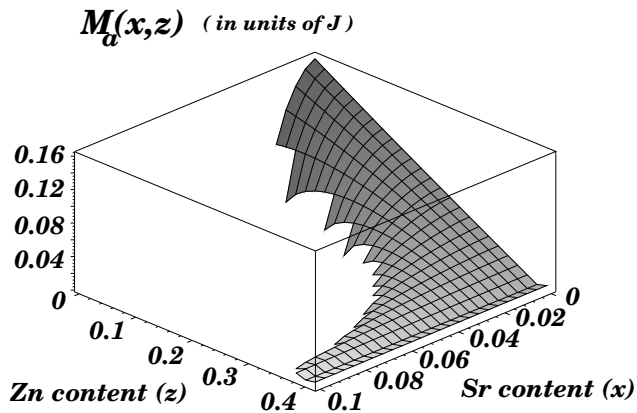}
\caption{3D plot with the dependence of the DM gap 
($\Delta_{DM}=M_a$, in units of $J$) on both Sr and Zn doping.}
\label{DM-3D}
\end{figure}
%

Let us now list a number of predictions from our studies, for a 
few selected doping concentrations of Sr and Zn, that can be 
verified experimentally. 

\begin{enumerate}

\item For La$_2$Cu$_{0.96}$Zn$_{0.04}$O$_4$, that is $x=0$ and 
$z=0.04$ where $\sqrt{Z}\approx 0.9$, we estimate that the 
reduction of the DM gap will be of about $10\%$ from the undoped 
$x=z=0$ value.  

\item For La$_{1.99}$Sr$_{0.01}$Cu$_{0.97}$Zn$_{0.03}$O$_4$, 
that is $x=0.01$ and $z=0.03$ where $\sqrt{Z}\approx 0.92$, 
the reduction of the DM gap will be smaller than the naive 
$8\%$, due to Zn, over the $30\%$ due to Sr. This is because 
of the dilution of frustration in the dipole-magnon coupling.

\item For La$_{1.98}$Sr$_{0.02}$Cu$_{0.85}$Zn$_{0.15}$O$_4$,
that is $x=0.02$ and $z=0.15$ where $\sqrt{Z}\approx 0.6$, we 
predic that the Dzyaloshinskii-Moriya gap is actually nonzero. 
In fact, due to the reduction of $40\%$ in the dipole-magnon
interaction, the DM gap is found to be $7.5$ cm$^{-1}$ and 
thus large enough, so that it can be accessed with one-magnon 
Raman spectroscopy.

\end{enumerate}

\subsection{XY or out-of-plane gap}

The mechanism for the reduction of the XY gap is the same as
the one discussed in the previous section, see Eq.\ (\ref{Mass-Ren}). 
For the XY gap the
available data also comes from one-magnon Raman spectroscopy 
experiments.\cite{Gozar} For $x=z=0$ and at $10$ K the XY gap is 
$m_c=36$ cm$^{-1}$ (or $4.3$ meV), and is reduced by almost 
$15\%$ at $x=1\%$. 

%
\begin{figure}[t]
\includegraphics[scale=0.35,angle=-90]{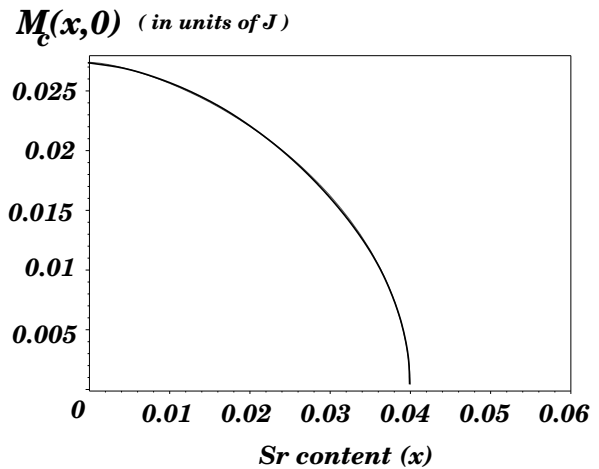}
\caption{Dependence of the XY gap ($\Delta_{XY}=M_c$, in units
of $J$) on Sr doping $x$ at zero Zn concentration: $z=0$.}
\label{XY-of-x}
\end{figure}
%

The dependence of the XY gap on Sr doping, according to our
theory, is shown in Fig.\ \ref{XY-of-x}. We find that at
$x=1\%$ and $z=0$ the XY gap is reduced by almost $10\%$, 
which is not far
from the experimentally measured value, and it vanishes
at $x=0.04$. When Zn is also included in the calculations
we obtain a similar nonmonotonic and reentrant behaviour, as 
observed for the DM gap (see the 3D plot in Fig.\ \ref{XY-3D}).

%
\begin{figure}[b]
\includegraphics[scale=0.35,angle=-90]{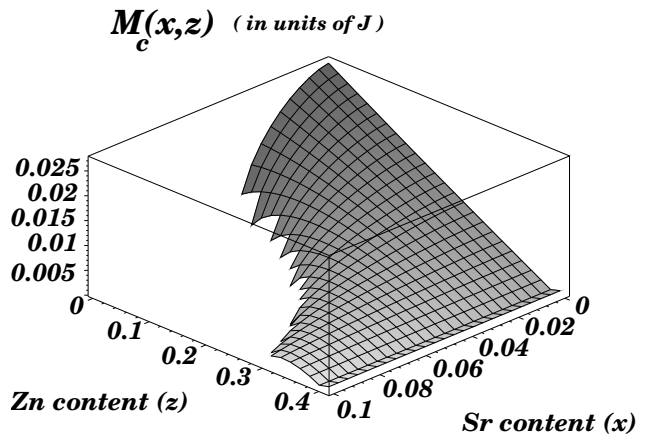}
\caption{3D plot with the dependence of the XY gap 
($\Delta_{XY}=M_c$, in units of $J$) on both Sr and Zn doping.}
\label{XY-3D}
\end{figure}
%

The vanishing of the XY gap at $x=4\%$ for $z=0$ has another
very interesting consequence. As it has been proposed by Juricic
{\it et al.},\cite{VladimirHeli} for $z=0$ and $x>0.02$, after 
the DM gap has disappeared, 
the magnetism becomes incommensurate (the staggered moment becomes
helicoidal along the $b$-axis) with an incommensurate wave vector 
${\bf Q}\parallel b$ and with magnitude
\beq
Q=\sqrt{x^2-\tilde{M}_c(x,0)^2},
\eeq
where $\tilde{M}_c=M_c/\rho_s$ is dimensionless. As a consequence of the 
nonzero character of $M_c$, the incommensurability, for 
$0.02<x<0.04$, deviates from the linear
behaviour $Q=x$, as indeed observed experimentally.\cite{Matsuda} 
For higher Sr doping, however, when $M_c(x>0.04)=0$, the linear 
relation $Q=x$ becomes exact.\cite{Matsuda}

\subsection{Order parameter}

Because of the DM and XY anisotropies, at zero applied magnetic 
field the staggered order parameter is oriented along the orthorhombic 
$b$ axis with magnitude $\sigma_0(x,z)$. In the presence of a 
longitudinal magnetic field, ${\bf B}\parallel b$, the spins 
start to rotate on the $bc$ plane as described theoretically by 
Silva Neto and Benfatto in Ref.\ \onlinecite{Marcello-Lara} 
(see also Benfatto {\it et al.} in Ref.\ \onlinecite{LaraRaman}) 
and experimentally measured with neutron diffraction by Reehuis 
{\it et al.} in Ref.\ \onlinecite{Reehuis}. At a critical field 
determined by the DM gap
\beq
H_c(x,z)=M_a(x,z),
\eeq
a spin-flop transition occurs, where the in-plane component of the order
parameter becomes oriented along the orthorhombic $a$ axis. Thus,
any nonmonotonic and reentrant behaviour of the DM gap will result
on a similar reentrant behaviour for the critical field. This
is a prediction that could be detected by neutron scattering experiments.

\subsection{N\'eel temperature}

In the case of single crystals doped with Sr and Zn, the available 
data from Ref.\ \onlinecite{Hucker} were obtained with samples which
contained less than 30$\%$ of Zn. Therefore, in this doping range
the dependence of $T_{N}(z)$ is linear and is well described with
the bond percolation factor $K(z)=1-3z$.

%
\begin{figure}[htb]
\includegraphics[scale=0.35,angle=-90]{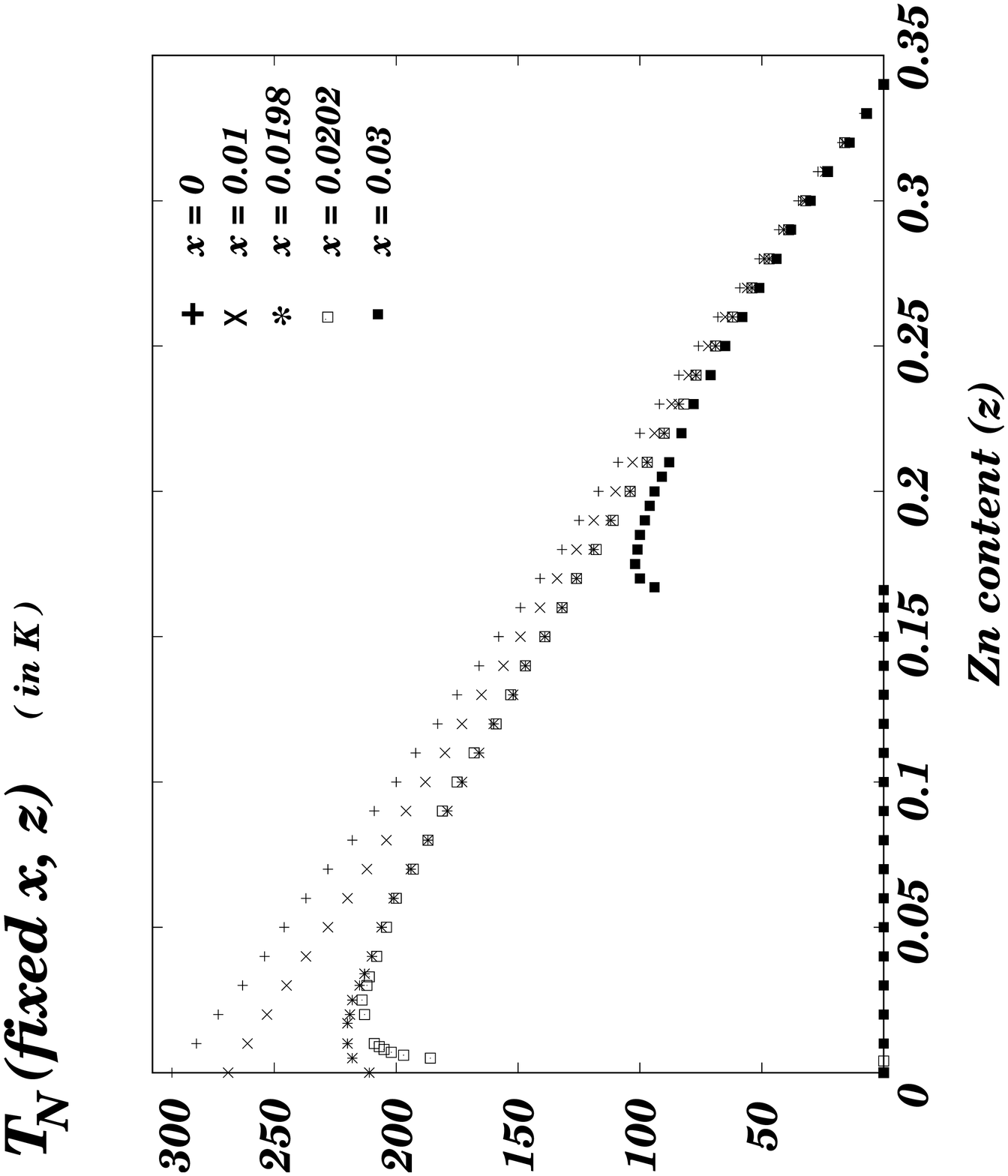}
\caption{Dependence of $T_{N}$ (in K) on Zn doping $z$ at various fixed 
Sr concentrations $x=0, 0.01, 0.0198, 0.0202, 0.03$. The 
nonmonotonic and reentrant behaviours are evident.}
\label{Fixed_Sr}
\end{figure}
%

The N\'eel temperature, $T_N(x,z)$, where the order parameter
vanishes, is given by Eq.\ (\ref{T-Neel}), with the stiffness and
magnon gaps renormalized by the impurities, see Eqs.\ (\ref{Mass-Ren})
and (\ref{Stiff-Ren}). As it happens with the
DM gap, we find that $T_N$ exhibits a nonmonotonic and reentrant
behaviour as the Zn concentration is increased, see Fig.\ \ref{Fixed_Sr}. 
In particular, we
find a monotonic decrease in the slope of the curves
$T_N(x,z=0)$ and $T_N(x,z=0.15)$, in qualitative agreement with the
experiments of H\"ucker {\it et al.}\cite{Hucker}, see 
Fig.\ \ref{Fixed_Zn}. Since we neglect
the self-consistent renormalization of the magnon gaps with
temperature, the agreement of our theoretical curves for $T_N$
with experiments is only qualitative, in contrast to the 
low-temperature results for the DM gap discussed above, which are
quantitative.

%
\begin{figure}[htb]
\includegraphics[scale=0.35,angle=-90]{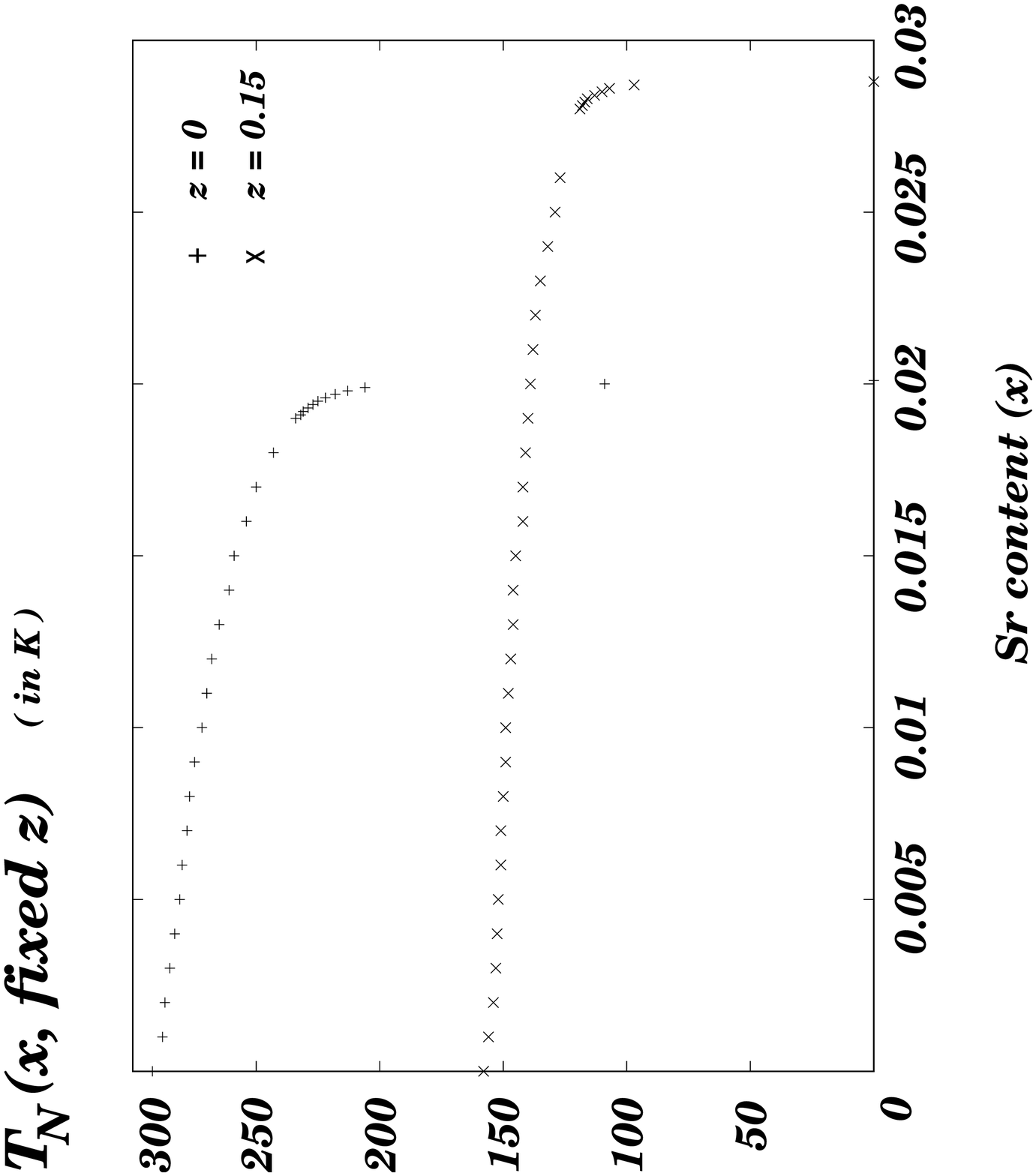}
\caption{Dependence of $T_{N}$ (in K) on Sr doping $x$ at two different
Zn concentrations: $z=0$ and $z=0.15$. The slope of the curves
decreases monotonically with increasing Zn content.}
\label{Fixed_Zn}
\end{figure}
%

It is worth emphasizing that the nonmonotonic behaviour exhibited
by our theoretical curve $x=0.0198$ in Fig.\ \ref{Fixed_Sr} is 
experimentally observed already at $x=0.017$. Moreover, the 
2D-Ising-like behaviour of the curves in Fig.\ \ref{Fixed_Zn} are 
actually an artifact of the dimensionality (we are considering an 
easy-axis 2D NLSM) and of the approximation (we are neglecting the 
thermal renormalizations of the gaps). We expect that by including 
the self-consistent thermal renormalizations and interlayer coupling, 
the agreement between theory and experiment in Fig.\ \ref{Fixed_Zn} 
will be satisfactory. Nevertheless, the qualitative agreement with
experiments gives strong support to the model and mechanisms 
considered here.

\subsection{The weak-ferromagnetic moment}

The staggered pattern of tilted octahedra in {\lco} is known to be
responsible for a weak-ferromagnetism, signaled by a cusp in
the low-field magnetic susceptibility.\cite{Thio} Within the language
of the NLSM (see Ref.\ \onlinecite{Marcello}), such weak-ferromagnetic
moment is proportional to the N\'eel order parameter via the
equation
\beq \langle{\bf L}\rangle=\frac{1}{2J}\left[\langle{\bf
n}\rangle\times{\bf D}_+\right].\eeq
Since within the N\'eel phase the average value of the staggered
magnetization is $\langle{\bf n}\rangle=(0,\sigma_0,0)$ (in the
$abc$ orthorhombic coordinate system), any nonmonotonic and
reentrant behaviour observed in $\sigma_0$ will cause also a
similar effect in the weak-ferromagnetic moment, and this can be
accessed in magnetic susceptibility experiments.

\section{Conclusions and outlook}

\label{Conclusions}

In this paper we revisited the problem of the dilution of
frustration in {\lasczo}, within the framework of a generalized
NLSM that includes DM and XY anisotropies. We showed that dilution
by Zn weakens the frustration by Sr through the reduction of the
dipole-magnon coupling constant, see Eq.\ (\ref{Klambda}). This 
leads to a nonmonotonic and reentrant behavior not only for 
$T_N$ but also for other observables like the order parameter, 
the weak-ferromagnetic moment, and the anisotropy gaps. 

Most remarkably, we predict that for $x\approx 2\%$ and $z=15\%$ 
the DM gap is approximately $7.5$ cm$^{-1}$, that is, larger than 
the lowest low-frequency cutoff for Raman spectroscopy ($\sim 5$ 
cm$^{-1}$) and thus likely to be observed in one-magnon Raman 
scattering. Furthermore, when the WL expression for the bond 
percolation factor is incorporated into our NLSM description, not 
only as a reduction factor for the spin-stiffness but also and 
most importantly for the reduction of the anisotropy gaps, we 
find that our NLSM with dilution describes correctly the data 
for $T_N(x=0,z)$, also in the highly Zn-diluted regime. Finally, 
we have also found that the XY gap vanishes, in the absence of 
dilution, for $x=0.04$ and this is consistent with the deviation 
from linearity, for $0.02<x<0.04$, of the incommensurate peaks 
seen in neutron scattering within the spin-glass phase of {\lasco}.

\section{Acknowledgements}

The authors acknowledge useful discussions with A. Aharony, 
Y. Ando, G. Blumberg, Yu-Chang Chen, and Nils Hasselmann.


\appendix

\section{Derivation of site and bond percolation factors}

\label{Appendix-A}

This derivation can be found also in Ref. \onlinecite{ChenThesis}, but
for the sake of completeness we will here go through the basic
steps of the derivation.

Averaging the site percolation factor we obtain
\begin{eqnarray}
P_{\infty}&=& P_{\infty}(z)=\frac{1}{Na^{2}} \int
d^{2}{\bf r} p({\bf r})=\frac{a^2(N-N_{Zn})}{Na^{2}} \nonumber \\
&=& 1-N_{Zn}/N=1-z,
\end{eqnarray}
where averaging $p({\bf r})$ over a 2D volume implies that we have
to normalize it with $Na^{2}$, $N$ being the number of sites in the
Cu-lattice.

We consider $p_{j}$ to be smooth, and thus we can expand it in
the neighborhood of the $i$-th site $p_{j}=p_{i}+{\bf a\nabla}p_{i}$.
Hence, the bond percolation factor $K_{ij}=p_{i}p_{j}$ in the
continuum limit becomes
$$
K({\bf r})=p({\bf r})[p({\bf r})+{\bf a\nabla}p({\bf r})]=p({\bf
r})+ \frac{1}{2} {\bf a\nabla}p({\bf r}),
$$
leading to
\begin{eqnarray}
K&=&K(z)=\frac{1}{Na^{2}} \int d^{2}{\bf r} \Big[p({\bf r})+
\frac{1}{2} {\bf a\nabla}p({\bf r}) \Big] \nonumber \\
&=& 1-z+\frac{{\bf a}}{2Na^{2}} \int d^{2}{\bf r}
\nabla p({\bf r}) \nonumber \\
&=& 1-z+ \frac{{\bf a}}{2Na^{2}}(-1) \sum_{Zn}\int
\mbox{contours} \nonumber \\
&=& 1-z-\frac{{\bf a}N_{Zn}4{\bf a}}{2Na^{2}}=1-3z
\end{eqnarray}
Using Stoke's theorem in the above formula we obtained the
integration over a contour of the 2D volume, which splits into the
sum of contours over Zn impurities. The minus sign accounts for
the opposite orientation of the small contours with respect to the
larger one.

\section{NLSM with dilution}

\label{Appendix-B}

In {\lco} the DM vectors are in good approximation perpendicular 
to the Cu-Cu bonds and change sign from one bond to another, while
the XY matrices provide an easy-plane anisotropy. It is worth noting 
that the pure 2D system defined by the action (\ref{Action}) does 
not display a rotational symmetry, so it can have order at finite 
temperature without violating the Mermin-Wagner theorem.

Before we proceed with the derivation of the NLSM, let us recall some
technical details: ${\bf d}_{+}$ is a vector in the $a$ orthorhombic 
direction; and $i$ is an index on a 2D lattice, so $i$ should be 
understood as $(p,q)$ and $(-1)^{i}$ as $(-1)^{p+q}$. Furthermore, 
we shall use that
\beq 
\sum_{<i,j>}{\bf D}_{ij}=\sum_{i} {\bf D}_{i,x}+
{\bf D}_{i,y}=2\sum_{i}{\bf d}_{+,i}, 
\eeq
\beq 
\oi=(-1)^{i}\nib+a{\bf l}_{i}-
\frac{1}{2}(-1)^{i}a^{2}\nib{\bf l}_{i}^{2},
\label{B4}
\eeq
with $\nib\cdot\li=0$,
\beq 
\njb=\nib - r_{ij}^{l}\partial_{l}\nib
+\frac{1}{2}r_{ij}^{l}r_{ij}^{m}\partial_{l}\partial_{m}\nib, 
\label{B5}
\eeq
and finally
\beq a^2 \sum_{i}=\idr. 
\label{nlsm3} 
\eeq

Let us first transform the Heisenberg term of the Hamiltonian. 
Using Eqs. (\ref{B4}) and (\ref{B5}) and neglecting $O(a^{3})$ 
terms we get
\bea \nonumber H_{H}&=& JS^{2}\sum_{<i,j>}\oi
\oj\nonumber\\
&=&JS^{2}\sum_{<i,j>}\left[
\frac{1}{2}r_{ij}^{l}r_{ij}^{m}\partial_{l}\nib\partial_{m}\nib
+2a^{2}\li^{2} \right]\nonumber\\&=& JS^{2}a^{2}\sum_{i}\left[
\frac{1}{2}(\nabla \nib)^{2}+4\li^{2} \right]. 
\label{nlsm7} \eea
In the continuum limit the Heisenberg Hamiltonian reads
\beq H_{H}=\frac{JS^{2}}{2}\idr\left[ (\nabla \nb)^{2}+8\lb^{2}
\right]. \label{nlsm8} \eeq
In the diluted case, we should multiply $\oi$  with $p_{i}$.
Hence, Eq. (\ref{nlsm7}) would be modified as follows
$$ H_{H}^{d}=JS^{2}\sum_{<i,j>}p_{i}p_{j}\oi\oj ,$$ where the index "d"
stands for "diluted". We want to treat the {\it static impurities}
as an {\it average effect}, thus we substitute $p_{i}p_{j}$ in the
above equation with $<p_{i}p_{j}>=K({\bf r})$, the bond
percolation factor (for more details, see chapter 3 or
Ref.\ \onlinecite{ChenThesis}). We simplify the problem even more by
considering the averaged one, taking $<K({\bf
r})>=K(z)$ (for the shortening of the notations in what
follows we will use K instead of K(z)), where $z$ is the Zn
concentration. Thus, in the diluted case  Eq.\ (\ref{nlsm8}) reads
\beq H_{H}^{d}=\frac{JS^{2}K}{2}\idr \left[ (\nabla
\nb)^{2}+8\lb^{2} \right] \label{nlsm9}. \eeq
Following a similar procedure, we may transform the DM and XY
terms of the Hamiltonian. For the DM Hamiltonian we get
\beq H_{DM}^{d}=\frac{4S^{2}K}{a}\idr \left[ {\bf
d}_{+}\cdot(\nb\times\lb) \right]. 
\label{nlsmDM} 
\eeq
For the XY Hamiltonian in the continuum limit we find
\beq H_{XY}^{d}=\frac{2S^{2}K}{a^{2}}\idr\left[
(\Gamma_{1}-\Gamma_{3})n_{z}^{2} \right], \label{nlsmXY} \eeq
where we neglected the small terms like $\Gamma(\nabla \nb)^{2}$
and $\Gamma \lb^{2}$.

Now, we will discuss in detail a Wess-Zumino term in 2D for the
diluted case, since it didn't appear in the literature. For the 1D
case we refer the reader to Ref.\  \onlinecite{Fradkin} 
(clean system) and Ref.\ \onlinecite{ChenThesis} (diluted system).

Following Fradkin (see Appendix A or Ref.\ \onlinecite{Fradkin}), we
write the Wess-Zumino action on a lattice (notice that
$(p,q)$ are the indices along x and y directions respectively, in
contrast with $i$ and $j$, which take values on a 2D lattice)
\beq 
\delta S_{WZ}=\delta\ob\cdot\ob\times\partial_{0}\ob 
\label{B12}
\eeq
\bea S_{WZ}&=& S\int_{0}^{T}dx_{0}\sum_{p,q}S_{WZ}[\ob(p,q)]
\nonumber \\ &=&
\frac{S}{2}\int_{0}^{T}\left[\sum_{p=1}^{N_{x}/2}\sum_{q=1}^{N_{y}}
\left\{S[\ob(2p,q)]+S[\ob(2p-1,q)]\right\} \right. \nonumber \\
&+& \left.
\sum_{p=1}^{N_{x}}\sum_{q=1}^{N_{y}/2}\left\{S[\ob(p,2q)]
+S[\ob(p,2q-1)]\right\}\right]
\label{B13}
\eea
In the second and third lines of Eq. (\ref{B13}) we recognize 
$\delta S_{WZ}$, which can be expressed through $\ob$. Using 
the spin decomposition (\ref{B4}), we get
\bea 
\delta_{x}\ob(p,q)&=&\ob(2p,q)+\ob(2p-1,q)\nonumber\\
\nonumber &=&(-1)^{2p}[\nb(2p,q)-\nb(2p-1,q)]\nonumber\\
&+&2a\lb(2p,q)+O(a^{2})\nonumber \\ 
&=& a\partial_{x}\nb+2a\lb. 
\label{B14}
\eea
Analogously, we transform the third line of Eq. (\ref{B13})
\beq \delta_{y}\ob(p,q)=a\partial_{y}\nb+2a\lb. 
\label{B15}
\eeq
Let us note that
\beq
\sum_{p=1}^{N_{x}/2}\sum_{q=1}^{N_{y}}=\frac{1}{2}\sum_{\mbox{all
sites}}=\frac{1}{2}\sum_{i}\rightarrow\frac{1}{2a^2}\idr.
\label{B16}
\eeq

Plugging (\ref{B12}), (\ref{B14}) and (\ref{B15}) into (\ref{B13}) 
and having in mind (\ref{B16}), we obtain for the Wess-Zumino action 
(decomposing (\ref{B12}) into staggered and uniform components, we 
keep only first order terms in $a$)
\beq 
S_{WZ}=\frac{Sa}{4}\int_{0}^{T}
dx_{0}\sum_{i}[(\nabla\nib+4\li)\cdot(\nib\times\partial_{0}\nib)].
\label{nlsm17} 
\eeq
The first term in Eq.\ (\ref{nlsm17}) is a topological term. It was 
demonstrated by Haldane\cite{Haldane} that in D$>$1 this 
term sums to zero in the AF background.

The total Euclidean action reads
\beq S_{E}=-iS_{WZ}+S_{H}+S_{DM}+S_{XY} .\eeq
Now we are ready to write a Wess-Zumino Lagrangian
\beq 
L_{WZ}=-i\frac{S}{a}\idr [
\lb\cdot\nb\times\partial_{\tau}\nb]. 
\label{B19}
\eeq
In the presence of {\it dilution} the action (\ref{B19}) will be
multiplied with $P_{\infty}$. The explanation is the following: in
the expression for $\delta S_{WZ}$ (\ref{B12}) we will have a factor
$p^{3}$, which by definition is equal to $p$ (and we further
simplify the problem by taking $<p>=P_{\infty}$). We can carefully
do the procedure in Eq.\ (\ref{B13}) in the diluted case, but since we
neglect $O(a^{2})$ terms, the answer will be the same.

Let us now summarize the results obtained so far. The total Euclidean
action in the diluted system reads
\beq 
S=\idt\; L_{tot},
\label{B20}
\eeq
with $L_{tot}= L_{WZ}+L_{H}+L_{DM}+L_{XY}$ and
\bea && L_{WZ}= \frac{-i S \pinf}{a}\idr \left[
\lb\cdot(\nb\times\partial_{\tau}\nb) \right] ,\\ \nonumber &&
L_{H}=\frac{JS^{2}K}{2}\idr\left[ (\nabla \nb)^{2}+8\lb^{2}
\right],
\\ \nonumber &&L_{DM}=\frac{4S^{2}K}{a}\idr \left[ {\bf
d}_{+}\cdot(\nb\times\lb) \right] ,\\ \nonumber &&L_{XY}=
\frac{2S^{2}K}{a^{2}}\idr \left[ (\Gamma_{1}-\Gamma_{3})n_{z}^{2}
\right]. \label{nlsm21} \eea
Now we integrate out $\lb$ in a sense of a saddle point solution.
We have to find a solution of an equation $$\frac{\delta
L_{tot}}{\delta \lb}=0$$ and plug it into Eq.\ (\ref{B20}). Doing this,
we get
\beq \lb=\frac{i \pinf}{8JSaK}(\nb\times\partial_{\tau}\nb)+
\frac{1}{2Ja}(\nb\times{\bf d}_{+}), 
\eeq
and
\bea 
S&=&\frac{1}{2gc}\idt\idr \left[
\frac{\pinf}{K}(\partial_{\tau} \nb)^2+Kc^{2}(\nabla
\nb)^{2}\right.\nonumber\\
&+&\left. K{\bf D}_{+}^{2}n_{a}^{2}+K\Gamma_{c}n_{c}^{2} \right],
\label{B23}
\eea
where we defined
\bea \nonumber && gc=8Ja^{2} \mbox{ - (bare) inverse transverse susceptibility},\\
\nonumber && c=2\sqrt{2}J S a \mbox{ - (bare) spin-wave velocity},\\
\nonumber && {\bf D}_{+}=\sqrt{2gcS^{2}/Ja^{2}}{\bf
d}_{+}=2\sqrt{2}Sd\overrightarrow{e}_{a} \mbox{ - DM vector},\\
\nonumber && \Gamma_{c}=(4gcS^{2}/a^{2})(\Gamma_{1}-\Gamma_{3})
\mbox{ - XY anisotropy} .\eea
It is convenient to introduce the notations: $D_{+}=m_{a}$ and
$\Gamma_{c}=m_{c}^{2}$. Plugging this into Eq.\ (\ref{B23}) and
rewriting it in a conventional way we get a final expression for
the total action in the presence of dilution
\begin{widetext}
\beq S=\frac{1}{2gcK/\pinf}\idt\idr \left[ (\partial_{\tau}
\nb)^{2}+Z\{c^{2}(\nabla \nb)^{2}+m_{a}^{2}n_{a}^{2}+
m_{c}^{2}n_{c}^{2}\} \right],\eeq
\end{widetext}
where $Z=K^{2}/\pinf$.


\end{document}